\documentclass[11pt]{article}
\newcommand{\Comment}[1]{{}}
\usepackage{amsfonts,amsthm,amsmath,amssymb,slashed}
\usepackage[textwidth = 430 pt, textheight = 630 pt]{geometry}
\usepackage{color}

\Comment{\usepackage{color}
\definecolor{MyDarkBlue}{rgb}{0.15,0.15,0.45}
\usepackage[linktocpage=true]{hyperref}
\hypersetup{
colorlinks=true,
citecolor=MyDarkBlue,
linkcolor=MyDarkBlue,
urlcolor=MyDarkBlue,
pdfauthor={Jeff Murugan, Horatiu Nastase, Nitin Rughoonauth and Jonathan P. Shock},
pdftitle={Particle-vortex and Maxwell duality in the $AdS_4\times {\mathbb CP}^3$/ABJM correspondence},
pdfsubject={hep-th}
}

\usepackage[numbers,sort&compress]{natbib}
\usepackage{hypernat}}
\usepackage{graphicx}
\usepackage{cite}

\newcommand\ignore[1]{}
\def\one{{\,\hbox{1\kern-.8mm l}}}

\def\Tr{{\rm Tr\, }}

\def\a{\alpha}\def\b{\beta}

\def\d{\partial}

\def\Tr{\mathop{\rm Tr}\nolimits}

\newcommand{\Cset}{{\,\,{{{^{_{\pmb{\mid}}}}\kern-.45em{\mathrm C}}}}}

\newcommand{\be}{\begin{equation}}
\newcommand{\bea}{\begin{eqnarray}}

\newcommand{\ee}{\end{equation}}
\newcommand{\eea}{\end{eqnarray}}

\begin{document}

\renewcommand{\thefootnote}{\fnsymbol{footnote}}

\makeatletter
\@addtoreset{equation}{section}
\makeatother
\renewcommand{\theequation}{\thesection.\arabic{equation}}

\rightline{}
\rightline{}

\begin{flushright}
QGASLAB-14-01
\end{flushright}

\vspace{10pt}


\begin{center}
{\LARGE \bf{\sc Particle-vortex and Maxwell duality in the $AdS_4\times \mathbb{CP}^3$/ABJM correspondence}}
\end{center} 
 \vspace{1truecm}
\thispagestyle{empty} \centerline{
{\large \bf {\sc Jeff Murugan${}^{a,}$}}\footnote{E-mail address: \Comment{\href{mailto:jeff@nassp.uct.ac.za}}{\tt jeff@nassp.uct.ac.za}},
{\large \bf {\sc Horatiu Nastase${}^{b,}$}}\footnote{E-mail address: \Comment{\href{mailto:nastase@ift.unesp.br}}{\tt nastase@ift.unesp.br}}, 
{\large \bf {\sc Nitin Rughoonauth${}^{a,}$}}\footnote{E-mail address: \Comment{\href{mailto:nitincr@gmail.com}}{\tt nitincr@gmail.com}}
{\bf{\sc and}}
{\large \bf {\sc Jonathan P. Shock${}^{a,}$}}\footnote{E-mail address: \Comment{\href{mailto:jon.shock@gmail.com}}{\tt jon.shock@gmail.com}}
                                                           }

\vspace{.5cm}

 
\centerline{{\it ${}^a$
The Laboratory for Quantum Gravity \& Strings, }} \centerline{{\it
Department of Mathematics and Applied Mathematics, }} \centerline{{\it
University of Cape Town, Private Bag, Rondebosch, 7700, South Africa}}

\centerline{{\it ${}^b$ 
Instituto de F\'{i}sica Te\'{o}rica, UNESP-Universidade Estadual Paulista}} \centerline{{\it 
R. Dr. Bento T. Ferraz 271, Bl. II, Sao Paulo 01140-070, SP, Brazil}}


\vspace{1truecm}

\thispagestyle{empty}

\centerline{\sc Abstract}

\vspace{.4truecm}

\begin{center}
\begin{minipage}[c]{380pt}
{\noindent We revisit the notion of particle-vortex duality in abelian theories of complex scalar fields coupled to 
gauge fields, formulating the duality as a transformation at the level of the path integral. This transformation is then made symmetric and cast as a self-duality that maps the original theory into itself with the role of particles and vortices interchanged. After defining the transformation for a pure Chern-Simons gauge theory, we show how to embed it into (a sector of) the $(2+1)-$dimensional ABJM model, and argue that this duality can be understood as being related to 4-dimensional Maxwell duality in the $AdS_{4}\times\mathbb{CP}^{3}$ bulk.
}
\end{minipage}
\end{center}

\vspace{.5cm}

\setcounter{page}{0}
\setcounter{tocdepth}{2}

\newpage

\renewcommand{\thefootnote}{\arabic{footnote}}
\setcounter{footnote}{0}

\linespread{1.1}
\parskip 4pt



\section{Introduction}
\ \ \ \ \ 
Non-perturbative dualities in quantum field theories have delivered many profound insights over the past three or so decades. 
Most famous among these are the lessons that we have learned about the very nature of spacetime via the duality between strongly 
coupled quantum field theories and theories of gravity as manifested in the AdS/CFT correspondence \cite{Maldacena:1997re}. 
Within the realm of quantum field theories alone, non-perturbative dualities rely on the fact that the generating functions 
of observables include an integration over the degrees of freedom. Consequently,  the choice of degrees of freedom with which 
we describe the system may result in multiple possibilities. In four dimensions, for example, the electromagnetic duality, 
manifest in the Maxwell equations, allows us to describe a system in terms of electric or magnetic fields and charges and exchanges 
fundamental particles for solitonic degrees of freedom. We therefore have a choice as to how we describe the system, and at the 
perturbative level, one or the other may be more appropriate depending on the problem at hand. This electric-magnetic duality 
(and its extension by Witten and Olive \cite{Witten:1978mh}) has had a powerful impact, not only on our understanding of the structure of gauge theories, but also on some of the deepest mathematical puzzles of our time \cite{Kapustin:2006pk}.  

A 3-dimensional analogue of the 4-dimensional duality above is one which exchanges fundamental particles with solitonic vortices but, defined only for abelian gauge theories, this particle-vortex duality, as well as  its physical implications, is much less understood than its 4-dimensional counterparts. 
To set the scene for what follows, we will first give an heuristic description of the particle-vortex duality elaborating on a 
discussion in the textbook of Zee \cite{Zee}, before embarking on a more technical treatment in the following section. Like many concepts commonplace in high energy theory, particle-vortex duality has its roots in the landscape of condensed matter; in this case in the theory of anyonic superconductivity \cite{Lee:1991jt}. After some limited further development in condensed matter physics, it was in the context of string theory that more development occured, 
starting with Intriligator and Seiberg \cite{Intriligator:1996ex}. Following 
\cite{Zee} then, we start with an abelian Higgs model 
\be
{\cal L}=-\frac{1}{2}|(\d_\mu -iq A_\mu)\phi|^2-V(\phi^\dagger \phi),
\ee
with some well-behaved potential, $V(\phi^{\dagger}\phi)$ for the complex scalar $\phi$.
We will ignore the potential term from now on, but presume that the theory exhibits vortex solutions (and consequently restrict our attention to three dimensions). Writing $\phi=|\phi|e^{i\theta}$ and restricting to the solution for which $|\phi|=v$ minimizes the potential gives 
\be
{\cal L}=-\frac{1}{2}v^2(\d_\mu \theta -qA_\mu)^2\label{thetaaction}.
\ee
We can introduce a non-dynamical field $\xi_\mu$ and write the Lagrangian in first order form as 
\be
{\cal L}=+\frac{1}{2v^2}\xi_\mu^2-\xi^\mu(\d_\mu \theta -qA_\mu)\,,
\ee
The phase $\theta$, characterizing the vortex is, in fact, 
singular at the origin for a vortex solution, allowing us to split it into a smooth part, and a vortex part:
\be
\theta=\theta_{smooth}+\theta_{vortex}\,,
\ee
where the vortex monodromy  $\Delta \theta_{vortex}=2\pi$. Integrating out $\theta_{smooth}$ gives $\d_\mu \xi^\mu=0$ and implies that we can write $\xi^\mu$ as the curl of a vector field
\be
\xi^\mu=\epsilon^{\mu\nu\rho}\d_\nu a_\rho\,.
\ee
Having integrated out $\theta_{smooth}$ and substituted in the new expression for 
$\xi^\mu$, we get the following Lagrangian
\be
{\cal L}=-\frac{1}{4v^2}f_{\mu\nu}^2+\epsilon^{\mu\nu\rho}\d_\nu a_\rho(\d_\mu \theta_{\rm vortex}-qA_\mu)\, ,
\ee
where $f_{\mu\nu}$ is the field strength tensor for $a_{\mu}$. A subsequent integration by parts and rewriting of the resulting term as
\be \label{eq.jvort}
a_\rho\epsilon^{\rho\mu\nu}\d_\mu \d_\nu \theta_{\rm vortex}=2\pi a_\mu j^\mu_{\rm vortex}\, ,
\ee
finally gives
\be
{\cal L}=-\frac{1}{4v^2}f_{\mu\nu}^2+2\pi a_\mu j^\mu_{vortex}-A_\mu J^{\mu},
\ee
where $J^\mu = q\epsilon^{\mu\nu\rho}\partial_\nu a_\rho$. Note that equation (\ref{eq.jvort}) is a crucial step in this derivation. Since 
the derivatives are contracted with the epsilon tensor, by symmetry arguments this expression will naively vanish. 
The only time this will not be the case is when there are singularities in $\theta$. Thus, $\theta_{vortex}$ is explicitly 
that part of $\theta$ which is not smooth and whose second derivative is related through this equation to the vortex current. {\it If there are no vortices then this expression will vanish and there will be no duality}. 

The introduction of an auxiliary vector field $\xi_\mu$ leads to the coupling of the vortex current to the gauge field $a_\mu$. We have gone from a description 
where the fundamental excitations are the particles associated to the field $\theta$ to the vortex description where the fundamental degrees of freedom are 
the vortices associated to $\theta_{vortex}$. However, to complete this description, we must have a field whose fundamental excitations themselves are vortices. 
To that end, we introduce a new field $\Phi$ which couples to $a_\mu$ precisely for this purpose. On adding this field, we can define an action which
gives a dual description, with particle and vortex degrees of freedom swapped. The Lagrangian
\be
{\cal L} = -\frac{1}{4v^2}f_{\mu\nu}^2 - \frac{1}{2}|(\d_\mu -i 2\pi a_\mu)\Phi|^2 - V(|\Phi|^2) - A_\mu(q\epsilon^{\mu\nu\rho}\d_\nu a_\rho)\, ,
\ee
then describes an abelian Higgs model for the vortex field $\Phi$ coupled to $a_\mu$ as opposed to the original field where $\theta$ was coupled to $A_\mu$.
The action of the transformation
\be
\d_\mu \theta -q A_\mu =\xi_\mu=\epsilon_{\mu\nu\rho}\d^\nu a^\rho\, ,\label{duality1}
\ee
exchanges the scalar degree of freedom $\theta$ with the gauge field degree of freedom $a_\mu$ in the presence of the background gauge field $A_\mu$. However, the necessity of introducing the new field $\Phi$ does not feel very satisfactory. We will see that there is a more complete way to formalise the duality\footnote{A more precise definition of particle-vortex duality, and an undertanding of how it arises in a path integral formulation was given by Burgess and Dolan in \cite{Burgess:2000kj}. For completeness, we review their formulation in Appendix A.}. The above transformation is also not strictly true in the presence of $\Phi$. 

A supersymmetric generalization of these ideas was proposed in \cite{Kapustin:1999ha} (see also \cite{Sachdev:2010zz}), however a path 
integral transformation realizing the particle-vortex duality could only be reduced to an unproven identity. Witten \cite{Witten:2003ya} later defined an $Sl(2,{\mathbb Z})$ transformation on a conformal field theory by combining an $S$-transformation (which adds an $\epsilon B \d A$ term to the Lagrangian) with a $T$-transformation (which adds a Chern Simons term, $\epsilon A \d A$). For example, starting with a charged scalar Lagrangian of the form ${\tilde L}(\Phi,A)$, the $TS$-transformation maps 
\be
{\tilde L}(\Phi,A)\ \xrightarrow{TS}\ L(\Phi,A,B)={\tilde L}(\Phi,A)+\epsilon^{ijk}B_i\d_jA_k+\epsilon^{ijk}A_i\d_jA_k\, .
\ee
The current-current two-point function of this three-dimensional CFT is constrained by conformal symmetry to be of the form:
\be
\langle J_i(k) J_j(-k)\rangle =(\delta_{ij}k^2-k_ik_j)\frac{t}{2\pi\sqrt{k^2}}+\epsilon_{ijk}k_k \frac{w}{2\pi}\, ,
\ee
where $t$ and $w$ form a complex coupling $\tau=w+it$. The action of the $TS$-transformations of the $Sl(2,{\mathbb Z})$ group on the complex parameter $\tau$ is then $\tau\rightarrow (a\tau+b)/(c\tau+d)$. Because this is an action on a conformal field theory, we can ask what the action of the transformation is on the gravity dual of this theory via the AdS/CFT correspondence. In this case the transformation acts on a $U(1)$ gauge field with a Maxwell action plus a topological theta term.

A while later, the constraints imposed on correlators in gauge theories by the existence of a particle-vortex duality were analysed in \cite{Herzog:2007ij}. Note that when the theory is changed by the action of the duality, (i.e. the theory is not self-dual), the correlators are themselves transformed. The authors also analyzed the $AdS_4\times S^7$ gravity dual of the ${\cal N}=8$ three-dimensional $SU(N)$ SYM in the large $N$ limit, and found that Maxwell duality in the bulk leads to the same type of constraints on correlators as would be obtained from a self-dual field theory. In abelian models a similar relation was obtained, and a correspondence with $AdS_4\times S^7$ was proposed as an implicit relation coming from large $N$ non-abelian gauge theories.

Today, the ABJM model \cite{Aharony:2008ug} is understood as the correct description of the field theory living on M2-branes and is dual (in the appropriate limit) to a type IIA supergravity on $AdS_4\times {\mathbb CP}^3$. This begs the question as to whether the results of \cite{Herzog:2007ij} can be reinterpreted from this point of view\footnote{The ABJM theory is also known to admit a maximally supersymmetric mass deformation \cite{Gomis:2008vc,Terashima:2008sy}, which not only allows us to go away from the conformal limit but also contains a rich spectrum of solitonic excitations.}. 

The aim of this article is two-fold: first we seek to provide a more precise definition of the particle-vortex duality at the level of a path integral transformation then, using this, we attempt to embed the duality transformation in the ABJM model. 

The structure of the paper is as follows. In section 2 we revisit the formulation of the particle-vortex duality by retaining some features of the relation of \cite{Burgess:2000kj} (reviewed in Appendix A) and defining it as an action on the path integral of the theory. In particular we find that, by combining it with the Mukhi-Papageorgakis Higgs mechanism for three-dimensional Chern-Simons theories \cite{Mukhi:2008ux} (see also \cite{Chu:2010fk}), we can define it as a self-duality of abelian Chern-Simons theories. In section 3 we look explicitly at vortex solutions and the conditions under which they exist in such theories. 

In section 4, we embed the particle-vortex duality in ABJM, showing that the abelian duality is part of the (large $N$) non-abelian theory. Finally, in section 5, we show that the particle-vortex duality is naturally obtained as the boundary relation corresponding to Maxwell duality in the bulk, using the AdS/CFT prescription. Thus, as in \cite{Mohammed:2012gi,Mohammed:2012rd}, we see that by using an abelian reduction of ABJM to an interesting non-conformal theory, we learn something about the structure of ABJM.

\section{Abelian particle-vortex duality in the path integral}

In this section we will extend the path-integral formulation of \cite{Burgess:2000kj} to give a better definition of the particle-vortex duality in abelian theories. To this end, let us consider a path integral for an abelian Higgs model consisting of a complex scalar field $\Phi=\Phi_0 e^{i\theta}$ coupled to a $U(1)$ gauge field $a_\mu$. Any kinetic term for the gauge field will be no more than a spectator for the transformation, as in the Burgess-Dolan formulation described in Appendix A and we will not include it in what follows. There will also be a potential term for $\Phi_0$, $V(\Phi_0^2)$, but this will also be a spectator so we will also choose to omit it now. When we want to explicitly discuss vortex solutions however, the potential will be important and will be included. As long as we are never integrating over $a_\mu$ or $\Phi_0$ we do not need to consider these last terms we have mentioned.

\noindent
The partition function for the theory is
\bea
Z &=& \int {\cal D}a_\mu {\cal D}\Phi_0{\cal D}\theta \exp \left\{-\frac{i}{2} \int d^3x\ |(\d_\mu - iea_\mu)\Phi|^2\right\}\cr
&=&\int {\cal D}a_\mu {\cal D}\Phi_0{\cal D}\theta \exp \left\{-\frac{i}{2} \int d^3x\ \left[(\d_\mu \Phi_0)^2 + (\d_\mu \theta_{\rm smooth} + \d_\mu \theta_{\rm vortex} + ea_\mu)^2\Phi_0^2\right]\right\}\, ,\cr
&&
\eea 
where, as in the previous section, we have split the $\theta$ field into a smooth part, and a topologically non-trivial and non-smooth vortex part. We define $\lambda_\mu = \d_\mu \theta$, after which we promote it to an independent variable in a first order formulation. $\lambda_\mu = \d_\mu\theta$ follows from the constraint $\epsilon^{\mu\nu\rho}\d_\nu\lambda_\rho=0$, which can be imposed via a Lagrange multiplier $b_\mu$, giving the path integral for the `master' action
\begin{align}
Z &= \int{\cal D}a_\mu {\cal D}\Phi_0{\cal D}b_\mu {\cal D}\lambda_\mu \exp\left\{-\frac{i}{2}\int d^3x\ [(\d_\mu\Phi_0)^2 + (\lambda_{\mu,{\rm smooth}} + \lambda_{\mu,{\rm vortex}} + ea_\mu)^2\Phi_0^2\right. \cr
&\left. \phantom{\int{\cal D}a_\mu {\cal D}\Phi_0{\cal D}b_\mu {\cal D}\lambda_\mu \exp-\frac{1}{2}\int d^3x}\hspace{25pt} + \frac{1}{e}\epsilon^{\mu\nu\rho} b_\mu \d_\nu \lambda_\rho] \right\}\, .
\end{align}
Integrating over $b_\mu$ returns us to the original formulation for the partition function, establishing the self-consistency of our procedure. If however we integrate over $\lambda_\mu$ first, we obtain 
the equation of motion 
\be
(\lambda_{\mu, {\rm smooth}} + \lambda_{\mu,{\rm vortex}} + ea_\mu)e\Phi_0^2 = -\epsilon^{\mu\nu\rho}\d_\nu b_\rho\,,\label{duality3}
\ee
which, on substitution back into the action produces the path integral for the {\it dual} action,
\begin{align}\nonumber
Z &= \int{\cal D}a_\mu {\cal D}\Phi_0{\cal D}b_\mu \exp\left\{-i\int d^3x\ \left[\frac{1}{4e^2\Phi_0^2}f^b_{\mu\nu}f^{b \mu\nu} + \epsilon^{\mu\nu\rho}b_\mu \d_\nu a_\rho - \frac{2\pi}{e}j^\mu_{\rm vortex}(t)b_\mu\right.\right.\\  
&\left.\left.\phantom{= \int{\cal D}a_\mu {\cal D}\Phi_0{\cal D}b_\mu \exp\ -\int d^3x}\hspace{15pt} + \frac{1}{2}(\d_\mu\Phi_0)^2\right]\right\}\,,
\end{align}
where $j^\mu_{\rm vortex}(t)$ is the vortex current in (\ref{vortexcurrent}), i.e., 
\be
j^\mu_{\rm vortex}(t) = \frac{1}{2\pi}\epsilon^{\mu\nu\rho}\d_\nu\d_\rho\theta 
= \frac{1}{2\pi}\epsilon^{\mu\nu\rho}\d_\nu\d_\rho\omega = \sum_a N_a\ \dot{y}_a^\mu\ \delta[x-y_a(t)]\,,
\ee
and is associated with the existence of vortex boundary conditions for $\theta$ in the original action with vortices positioned at $\vec{y}_a(t)$ in the two dimensional space (see equation (\ref{A5}) for a definition of $\omega$). In the dual action, it appears as an explicit
source term. Here, the sum is over all vortex positions labeled by the index $a$. Also note that, as in (\ref{scalarcurrent}), $j_\mu=e\Phi_0^2\d_\mu\theta$ is a scalar current, and we have then the duality relation between the vortex current and the scalar current:
\be
j^\mu_{\rm vortex}(t)=\frac{1}{2\pi e\Phi_0^2}\epsilon^{\mu\nu\rho}\d_\nu j_\rho\,.\label{currentduality}
\ee
Notice that here $\Phi_0$ has the interpretation of a coupling constant for the field $b_\mu$ dual to $\theta$, which itself becomes a dynamical Maxwell gauge field. In this sense this duality maps particles to vortices, justifying the name {\it particle-vortex duality}. 

\subsection{The Mukhi-Papageorgakis Higgs mechanism}

There is a striking similarity between the particle-vortex duality described here and a version of the Higgs mechanism for three-dimensional Chern-Simons theories discovered by Mukhi 
and Papageorgakis in \cite{Mukhi:2008ux} in the context of ABJM theories, but valid more generally (see also \cite{Chu:2010fk} for more details about its
implementation). The statement analogous to the usual Higgs mechanism statement that a massless gauge field eats a scalar and becomes massive, is now 
that a Chern-Simons gauge field (with no dynamical degrees of freedom) eats a scalar and becomes dynamical, i.e. of Maxwell (or Yang-Mills) form with one dynamical degree of freedom. 

The mechanism itself goes as follows. We start with an action for a complex scalar, $\Psi$, coupled to a Chern-Simons gauge field, 
$a_\mu$,
\be
  S = -\int d^3x\ \left[\frac{k}{2\pi} \epsilon^{\mu\nu\rho}a_\mu \d_\nu \tilde a_\rho + \frac{1}{2}|(\d_\mu -iea_\mu)\Psi|^2 
  + V(|\Psi|^2)
\right]\,,\label{cs}
\ee
with a vacuum solution $\Psi=b$. We can then expand the scalar degrees of freedom around the ground state
\be
  \Psi = (b + \delta \psi)e^{-i\delta\theta};\;\;\;\; \delta \theta = \theta_{\rm smooth} + \theta_{\rm vortex}\, ,
\ee
and plug it back in the action to find
\be
  S = -\int d^3x\ \left[\frac{k}{2\pi} \epsilon^{\mu\nu\rho}a_\mu \d_\nu \tilde a_\rho+\frac{1}{2}(\d_\mu \delta \psi)^2 +   
  \frac{1}{2}(\d_\mu \theta_{\rm smooth} + \d_\mu \theta_{\rm vortex} + ea_\mu)^2b^2 + \ldots\right]\,.
\ee
Here, the omitted terms come from the $\delta\psi$ self-interaction in $V(|\Psi|^2)$ and the $\delta\theta$-$\delta\psi$ interaction. Note that, for the purposes of making a comparison, we have allowed for the possibility that $\delta\theta$ contains a vortex piece $\theta_{\rm vortex}$. The mechanism by which the Chern-Simons vector eats the scalar and becomes a dynamical Maxwell vector happens through exactly the same redefinition as in the the usual Higgs mechanism. Here we write
\begin{equation}\label{eq.MPmech}
ea_\mu + \d_\mu\theta_{\rm smooth} + \d_\mu \theta_{\rm vortex} = ea'_\mu\,,
\end{equation} 
trivially integrate out $\theta$ and add a boundary term to the action to obtain
\be
  S = -\int d^3x\ \left[\frac{k}{2\pi} \epsilon^{\mu\nu\rho}a'_\mu \d_\nu \tilde a_\rho + \frac{1}{2}(\d_\mu \delta \psi)^2 +   
  \frac{1}{2}(ea'_\mu)^2b^2
  - \frac{k}{e}  j^\mu_{\rm vortex}\tilde a_\mu + \ldots\right]\,.
\ee
Solving for $a'_\mu$ gives
\be
  a^\mu+\frac{1}{e}\d^\mu\delta \theta = a'^\mu = -\frac{k}{2\pi b^2}\epsilon^{\mu\nu\rho}\d_\nu \tilde a_\rho\,,
\ee
which is similar to (\ref{duality3}) for the particle-vortex duality. Defining $\tilde f_{\mu\nu}=\d_\mu \tilde a_\nu -\d_\nu\tilde a_\mu$, we find
\be
S = \int d^3x\ \left[-\frac{ k^2}{16\pi^2b^2}(\tilde f_{\mu\nu})^2 - \frac{1}{2}(\d_\mu \delta\psi)^2 + \frac{k}{e} j^{\mu}_{\rm vortex}\tilde a_\mu + \ldots \right]\,,\label{higgs}
\ee
where again some nonlinear terms in the fluctuations - specifically, the self-interactions of $\delta\psi$ coming from $V(|\Psi|^2)$ and terms that would appear when replacing $b$ in the Maxwell coupling with $|\Psi| = b + \delta\psi$ - are omitted with impunity, since they could be reintroduced by simply writing $|\Psi|$ instead of $b$ and retaining $V(|\Psi|^2)$. We close this section with two points of note. Firstly, the addition of a term $-\epsilon^{\mu\nu\rho}\tilde a_\mu \d_\nu b_\rho$ to either (\ref{cs}) or (\ref{higgs}) can be made without changing anything since the transformations don't act on either $\tilde a$ or $b$. Second, assuming vortex boundary conditions in the initial action give a vortex current coupling in the final action. Again, this is as in the case of 
particle-vortex duality, although here we can assume regular boundary conditions and thus avoid the vortex current $j^\mu_{\rm vortex}$.

\subsection{A symmetric duality}

As described in the previous section, particle-vortex duality is not a self-duality, in that it maps the original action to a manifestly different action. In particular it dualizes the scalar angle $\theta$ to the gauge field $b_\mu$. For our purposes of embedding the duality in the ABJM model, it will be useful to `symmetrize' this duality. As we demonstrate now, this may be acheived by adding a gauge field and a real scalar, and dualizing them to a complex scalar. This means that the original and final action will look the same.
As before, we may also add vortex currents. We will also omit a possible kinetic term for $a_\mu$ and explicitly write the self-interactions of the scalars $\Phi$ and $\chi$. Our launching point, again, will be the path integral
\begin{align}\nonumber
  Z &= \int{\cal D}a_\mu {\cal D}\Phi_0{\cal D}\chi_0{\cal D}\theta {\cal D}\tilde b_\mu \exp\left\{-i\int d^3x\ \left[\frac{1}  
  {2}|(\d_\mu - iea_\mu)\Phi_0e^{-i\theta}|^2 + \frac{1}{2}(\d_\mu \chi_0)^2\right.\right.\\
  &\phantom{\int\ \ \ } \left.\left. + \frac{1}{4e^2\chi_0^2}f_{\mu\nu}^{(\tilde{b})}f^{(\tilde{b})\mu\nu} + \epsilon^{\mu\nu  
  \rho}a_\mu \d_\nu \tilde b_\rho-\frac{2\pi}{e}\tilde b_\mu \tilde j^\mu_{\rm vortex}(t)
  +V(\Phi_0^2)+V(\chi_0^2)\right]\right\}\,,\label{symm1}
\end{align}
where $\tilde j^\mu_{\rm vortex}(t)$ is a source term that, in the dual version, will be associated to vortex boundary conditions for the dual scalar. $\tilde{b}_\mu$ is our new gauge field and $\chi_0$, the new scalar. It is the addition of these two that will lead to a self-dual action. We again write a first order formulation for $\lambda_\mu=\d_\mu\theta$ and then impose this relation as the constraint $\epsilon^{\mu\nu\rho}\d_\nu\lambda_\rho=0$ through a Lagrange multiplier $b_\mu$. Conversely, we can define $\tilde \lambda_\mu$ via a tilde version of (\ref{duality3}), namely
\be
(\tilde\lambda_{\mu,{\rm smooth}} + \tilde\lambda_{\mu,{\rm vortex}} + e a_\mu)e\chi_0^2 = -\epsilon^{\mu\nu\rho}\d_\nu\tilde b_\rho\,,
\ee
and then introduce $\tilde\lambda_\mu$ in the action such that we have the above equation as its equation of motion. Either way, we obtain the path integral for the master action
\bea
Z &=& \int {\cal D}a_\mu {\cal D}\Phi_0{\cal D}\chi_0{\cal D}\lambda_\mu {\cal D}b_\mu {\cal D}\tilde \lambda_\mu{\cal D}\tilde b_\mu \cr
&&\exp\left\{-i\int d^3x\ \left[\frac{1}{2}(\d_\mu\Phi_0)^2 + \frac{1}{2}(\d_\mu \chi_0)^2+ \frac{1}{e}\epsilon^{\mu\nu\rho}(b_\mu\d_\nu \lambda_\rho + \tilde b_\mu \d_\nu \tilde \lambda_\rho)\right.\right.\cr
&& \left.\left. + \frac{1}{2}(\lambda_\mu + \lambda_{\mu,{\rm vortex}} + ea_\mu)^2\Phi_0^2 + \frac{1}{2}(\tilde\lambda_\mu + \tilde \lambda_{\mu,{\rm vortex}} + e\tilde{a}_\mu)^2\chi_0^2 +V(\Phi_0^2)+V(\chi_0^2)\right]\right\}\,.\cr
&&
\eea
Now repeating the same procedure for the fields with tilde replaced with untilde (or, equivalently, integrating over $\lambda_\mu$ and $\tilde b_\mu$, to write $\tilde \lambda_\mu = \d_\mu \tilde \theta$), we obtain the path integral for the dual action
\begin{align}\nonumber
  Z &= \int{\cal D}a_\mu {\cal D}\Phi_0{\cal D}\chi_0{\cal D}\tilde\theta {\cal D} b_\mu \exp\left\{-i\int d^3x\ \left[\frac{1}  
  {2}|(\d_\mu -iea_\mu)\chi_0e^{-i\tilde\theta}|^2 + \frac{1}{2}(\d_\mu \Phi_0)^2\right.\right.\\
  &\ \ \ \left.\left. + \frac{1}{4e^2\Phi_0^2}f_{\mu\nu}^{(b)}  f^{(b)\mu\nu} + \epsilon^{\mu\nu\rho}a_\mu \d_\nu b_\rho -   
  \frac{2\pi}{e}b_\mu j^\mu_{\rm vortex}(t) + V(\Phi_0^2) + V(\chi_0^2)\right]\right\}\,.\label{symm2}
\end{align}
Assuming that $a_{\mu}$ has no kinetic term, we can now actually integrate it out in both the original and dual actions. Indeed, the terms containing $a_\mu$ in the Lagrangian (equation (\ref{symm1})) are 
\be
{\cal L}^{(a)}=-\frac{1}{2}e^2a_\mu^2|\Phi_0|^2-a^\mu(j_\mu+J_\mu)\, ,
\ee
with $j_\mu=-\frac{ie}{2}(\Phi\d_\mu \Phi^*-\Phi^*\d_\mu \Phi)=e\d_\mu\theta$, the scalar current and topological (vortex-like) current $J^\mu=\epsilon^{\mu\nu\rho}\d_\nu\tilde b_\rho$. Solving for $a_\mu$ we obtain
\be
a_\mu=-\frac{1}{e^2\Phi_0^2}(j_\mu +J_\mu)\, ,
\ee
and substituting back into ${\cal L}^{(a)}$, produces an extra contribution
\be
{\cal L}_{\rm extra}=+\frac{1}{2e^2\Phi_0^2}(j_\mu+J_\mu)^2=-\frac{1}{4e^2\Phi_0^2}\left(f_{\mu\nu}^{\tilde b}-\epsilon_{\mu\nu\rho}j^\rho\right)^2.
\ee
Having thus eliminated $a_{\mu}$ from the picture, we are now in a position to realize the duality as a map from 
\begin{align}\nonumber
Z &= \int{\cal D}\Phi_0{\cal D}\chi_0{\cal D}\theta {\cal D}\tilde b_\mu \exp\left\{-i \int d^3x\ \left[\frac{1}{2}|\d_\mu(\Phi_0e^{-i\theta})|^2 + \frac{1}{2}(\d_\mu\chi_0)^2 + \frac{1}{4e^2\chi_0^2} f_{\mu\nu}^{(\tilde b)} f^{(\tilde b) \mu\nu}\right.\right.\\ \nonumber
&\ \ \ \left.\left. + \frac{1}{4e^2\Phi_0^2}\left( f_{\mu\nu}^{(\tilde b)} - 
\epsilon_{\mu\nu\rho}j^\rho\right)^2 - \frac{2\pi}{e}\tilde b_\mu \tilde j^\mu_{\rm vortex}(t) + V(\Phi_0^2)+V(\chi_0^2)\right]\right\}\, ,\\
\end{align}
into 
\begin{align}\nonumber
Z &= \int{\cal D}\Phi_0{\cal D}\chi_0{\cal D}\tilde \theta {\cal D} b_\mu \exp\left\{-i \int d^3x\ \left[\frac{1}{2}|\d_\mu(\chi_0e^{-i\tilde\theta})|^2 + \frac{1}{2}(\d_\mu\Phi_0)^2 + \frac{1}{4e^2\Phi_0^2} f_{\mu\nu}^{(b)}  f^{(b) \mu\nu}\right.\right.\\ \nonumber
&\ \ \ \left.\left. + \frac{1}{4e^2\chi_0^2}\left( f_{\mu\nu}^{(b)} - 
\epsilon_{\mu\nu\rho}\tilde j^\rho\right)^2 - \frac{2\pi}{e}b_\mu j^\mu_{\rm vortex}(t) + V(\Phi_0^2) + V(\chi_0^2)\right]\right\}\, ,\\
\end{align}
that furnishes a formulation of the particle-vortex duality with an explicitly self-dual action.

Of course, since our aim is to embed the particle-vortex duality into the ABJM model and, in this case we have only scalars and a Chern-Simons gauge field at our disposal we will need to combine the symmetric form of the duality above with the Mukhi-Papageorgakis Higgs mechanism of the previous section. Moreover, in order for the duality to be nontrivial, we need to retain the vortex boundary conditions only in the original scalar, not the one that gets Higgsed. Starting from the path integral
\begin{align}\nonumber
Z &= \int{\cal D}a_\mu {\cal D}\Phi_0{\cal D}\theta {\cal D}\tilde b_\mu {\cal D}\chi {\cal D}\chi^*{\cal D}{\cal A}_\mu
\exp\left\{-i \int d^3x\ \left[\frac{1}{2}|(\d_\mu - iea_\mu)\Phi_0e^{-i\theta}|^2 \right.\right.\\
&\ \ \ \left.\left. + \frac{1}{2}|(\d_\mu - ie{\cal A}_\mu)\chi_0 e^{-i\phi}|^2 + \epsilon^{\mu\nu\rho}(\frac{1}{e}{\cal A}_\mu \d_\nu \tilde b_\rho + a_\mu \d_\nu\tilde b_\rho) + V(\phi_0^2) + V(\chi_0^2)\right]\right\}\,,\label{finalduality}
\end{align} 
we first implement the Mukhi-Papageorgakis Higgs mechanism by shifting ${\cal A}_\mu\rightarrow {\cal A}'_\mu$ as in equation (\ref{eq.MPmech}), absorbing $\phi$ and performing the (now trivial) path integral over $\phi$. Subsequently, we integrate over ${\cal A}'_\mu$ using the equation of motion
\be
e{\cal A}_\mu+\d_\mu \phi +\d_\mu \phi_{\rm vortex} \equiv e {\cal A}'_\mu =\frac{1}{e^2\chi_0^2}{\epsilon_\mu}^{\nu\rho}\d_\nu\tilde b_\rho\,,\label{calA}
\ee
and get exactly the path integral in (\ref{symm1}) which, as we saw previously, is dual to that in (\ref{symm2}). We now undo the Mukhi-Papageorgakis Higgs mechanism, by writing a first order formalism for $f_{\mu\nu}^{(b)}$ in terms of a field $\tilde {\cal A}'_\mu$, then introducing a trivial path integration over a variable $\tilde \phi$ and shifting ${\cal A}'_\mu$ by
\be
e\tilde {\cal A}_\mu + \d_\mu \tilde \phi + \d_\mu \tilde \phi_{\rm vortex} \equiv e \tilde{\cal A}'_\mu = \frac{1}{e^2\Phi_0^2}\epsilon^{\mu\nu\rho}\d_\nu b_\rho\, ,
\ee
so that we finally arrive at the path integral
\begin{align}\nonumber
Z &= \int{\cal D}a_\mu {\cal D}\chi_0{\cal D}\tilde\theta {\cal D}b_\mu {\cal D}\Phi {\cal D}\Phi^*{\cal D}\tilde {\cal A}_\mu
\exp\left\{-i \int d^3x\ \left[\frac{1}{2}|(\d_\mu - iea_\mu)\chi|^2 + \frac{1}{2}|(\d_\mu -ie\tilde {\cal A}_\mu )\Phi|^2\right.\right.\\
&\ \ \ \left.\left. + \epsilon^{\mu\nu\rho}(\frac{1}{e}\tilde{\cal A}_\mu \d_\nu b_\rho + a_\mu \d_\nu b_\rho) + V(\phi_0^2) + V(\chi_0^2)\right]\right\}\,,
\end{align}
where now $\chi=\chi_0e^{-i\tilde \theta}$ and $\Phi=\Phi_0 e^{-i\tilde\phi}$. Naively, it would seem that (\ref{calA}) undoes the duality transformation but it does not, since the interpretation is different. In the Higgs mechanism, we solve for ${\cal A}_\mu$ and $\phi$, while retaining $\tilde b_\mu$ in the theory. In the particle-vortex duality, we exchange $\tilde b_\mu$ for $\tilde \theta$ and similarly for quantities with tilde and untilde exchanged.

\section{Vortex solutions}

To summarize the story so far; we have formulated a manifest duality in the path integral formalism and argued that such a duality should exchange particles with vortices. Obviously, in order to do so, we need to have vortex solutions in the theory. Until now we have simply {\it presumed} the existence of such vortices in the field theories under investigation. Clearly this will not be the case for all field theories of the form we have been discussing. Here, therefore, we devote some time to discuss constraints on the form of the potential which will lead to such solutions. Thus, we consider the action in the path integral (\ref{finalduality}). In order to do this one first writes down the full equations of motion, and {\em only afterwards} will sets  $\chi=\tilde b_\mu ={\cal A}_\mu=0$ (which is itself a solution of these equations). The remaining equations of motion then become
\bea
&&\epsilon^{\mu\nu\rho}\d_\nu a_\rho=0\,,\cr
&& \Phi (D_\mu \Phi)^\dagger-\Phi^\dagger D_\mu \Phi=0\,,\label{eqs}
\eea
and the equation of motion for $\Phi$, which depends on the potential is
\be
D_\mu D^\mu \Phi=\frac{dV}{d|\Phi|^2}\,,\label{eqphi}
\ee
Note that the first of equations (\ref{eqs}) implies that $a_\mu$ must be pure gauge while the second equation means that 
\begin{equation}
D_\mu \theta=0\Rightarrow \d_\alpha \theta=a_\alpha\, , \label{eqalpha}
\end{equation}
where $\alpha$ is the polar angle in the complex plane, and $\theta$
is the argument of $\Phi$, i.e. $\Phi=|\Phi|e^{i\theta}$. In particular this relation is valid at infinity. This gives the usual charge quantization condition $\oint d\a\, a_\a=\oint d\theta$ which, in turn, implies that $\theta=N\a$. From the 0-component of equation (\ref{eqalpha}) we get for static solutions that $a_0=0$.

Note however that this result would imply as usual that $|\Phi(r=0)|=0$ for consistency of the vortex ansatz. This in turn means that the second equation in (\ref{eqs}) is already satisfied at $r=0$, hence we don't need $a_\a=N$ at $r=0$. That would be good, since substituting $a_\a=N$, requires that
\be
  \epsilon^{\mu\nu\rho}\d_\nu \d_\rho \theta\propto j^\mu_{\rm vortex}\propto \delta( r)\,,
\ee
so the first of equations (\ref{eqs}) would be satisfied everywhere except at $r=0$ which would in turn imply a discontinuous form for $a_\mu$ at $r=0$, necessitating some kind of regularization at this point. In fact, as we will soon demonstrate, in order to have a solution we need $|\Phi|\neq 0$ at $r=0$. Consequently, the solution as it stands will be valid everywhere except at $r=0$.
It remains now to satisfy the $|\Phi|$ equation of motion in order to determine the vortex profile. 
We know already from (\ref{eqphi}) that any vortex solution must satisfy 
\be
  \frac{|\Phi|''}{|\Phi|}=\frac{dV}{d|\Phi|^2}\,,\label{phir}
\ee
where, from general considerations about vortices, the one-vortex solution should behave like 
$|\Phi|\sim Ar$ as $r\rightarrow 0$. If in addition, we consider the most general renormalizable potential in three dimensions, namely the sextic, $V=C_1|\Phi|^6+\lambda |\Phi|^4+m^2|\Phi|^2$ for which $\frac{dV}{d|\Phi|^2}=m^2+2\lambda |\Phi|^2+3C_1|\Phi|^4$, 
several cases of interest for the asymptotic behaviour of these solutions present themselves. They are (in no particular order):

\begin{itemize}
\item
\underline{$m\neq 0$ and $\lambda\neq0$:} In this case, $V=C_1|\Phi|^6+\lambda |\Phi|^4+m^2|\Phi|^2$. Near the origin, we take as an ansatz for the field
\be
|\Phi|\sim Ar+Cr^p+...
\ee
This reduces the equation of motion in this region to
\be
\frac{p(p-1)Cr^{p-2}}{Ar}=m^2\,,
\ee
which fixes $p$ to be $3$ and $C=Am^2/6$. Therefore the small$-r$ form of the field is 
\be
|\Phi|\sim Ar\left(1+\frac{m^2}{6}r^2+...\right)\,.
\ee
Clearly we could go to any order analytically if needed. Taking the other asymptotic limit, if we chose that
as $r\rightarrow \infty$, $|\Phi|\sim \tilde A/r^n$, there is an inconsistency for non-zero $n$ as $|\Phi|''/|\Phi|\sim 1/r^2\rightarrow 0$, whereas $dV/d|\Phi|^2=m^2+...$. To avoid this, we choose instead
\be
|\Phi|\sim \tilde A+\frac{\tilde B}{r^n}+...
\ee
With this ansatz, the equation of motion reduces to
\be
\frac{\tilde B}{\tilde A}\frac{n(n+1)}{r^{n+2}}=(m^2+2\lambda \tilde A^2+3C_1\tilde A^4)+
\left(\frac{4\lambda \tilde A \tilde B+12 C_1\tilde A^3\tilde B}{r^n}\right)+
\frac{4\lambda \tilde B^2}{r^{2n}}+\frac{18 C_1\tilde A^2\tilde B^2}{r^{2n}}\,,
\ee
and we see that we need $n=2$ to satisfy the radial behaviour, along with the constraint that the two parentheses must vanish separately. 
From the first of these we find, 
\be
m^2+2\lambda \tilde A^2+3C_1\tilde A^4=0\,,
\ee
which says that $|\Phi|=\tilde A$ is the nontrivial vacuum of the theory, 
satisfying $dV/d|\Phi|^2=0$. The vanishing of the second parenthesis requires $\lambda +3C_1\tilde A^2=0$. Taken together, these two constraints give that,
\be
\tilde A^2=-\frac{m^2}{\lambda}\,,\,\,\,\,\,\, C_1=\frac{\lambda^2}{3m^2}\,.
\ee
Note that the latter is a constraint on the potential, allowing for only a certain class of sixth order potentials with non-zero quadratic and quartic terms to lead to vortex solutions. This tells us that $\lambda$ and thus $C_1$ need to be nonzero - ie. the potential must be truly sextic.
We then solve to the next order in $r$ in the equation of motion, i.e. $1/r^4$, giving
\be
\tilde B=\frac{3}{\tilde A(2\lambda +9C_1\tilde A^2)}\, ,
\ee
so that 
\be
|\Phi|\sim \tilde A+\frac{3}{\tilde A r^2(2\lambda +9C_1\tilde A^2)}+...
\ee
Clearly as $m\rightarrow 0$ this solution vanishes.

\item
\underline{$m=0$ and $\lambda \neq 0$:} In this case $V=C_1|\Phi|^6+\lambda |\Phi|^4$. As before we take the asymptotics close to the vortex origin to be
\be
|\Phi|\sim Ar+Cr^p+...
\ee
The equation of motion is now
\be
\frac{p(p-1)C r^{p-2}}{Ar}\simeq 2\lambda |\Phi|^2\simeq 2\lambda A^2r^2\, ,
\ee
which gives $p=5$ and $C=\lambda \frac{A^3}{10}$ so that
\be
|\Phi|\sim Ar\left(1+\frac{\lambda A^2}{10}r^4+...\right)\, .
\ee
Far away from the vortex we take $|\Phi|\sim \frac{\tilde A}{r^n}$ which reduces the equation of motion to
\be
\frac{n(n+1)}{r^2}\simeq 2\lambda |\Phi|^2=2\lambda \frac{\tilde A^2}{r^{2n}}\,.
\ee
This fixes $n=1$ and $\tilde A=1/\sqrt{\lambda}$, meaning that
\be
|\Phi|\sim \frac{1}{\sqrt{\lambda}r}+...
\ee
Note that  $\tilde A+\tilde B/r^n$ leads to a contradiction in the equations of motion and thus the leading term must be $\sim \frac{1}{r}$. In contrast to the first case above, there is no constraint on the potential. 

\item
\underline{$m=0$ and $\lambda=0$:} In this case $V=C_1|\Phi|^6$, a purely sextic potential). At $r=0$, as above, we find
\be
\frac{p(p-1)C r^{p-2}}{Ar}= 3C_1|\Phi|^4\sim 3 C_1 A^4r^4\, ,
\ee
which gives $p=7$ and $C=C_1 A^5/14$, so 
\be
|\Phi|\sim Ar\left(1+\frac{C_1A^4}{14}r^6+...\right)\,.
\ee
At infinity, with $|\Phi|\sim \tilde A/r^n$, the equation of motion is 
\be
\frac{n(n+1)}{r^2}=3C_1\frac{\tilde A^4}{r^{4n}}\, ,
\ee
which gives $n=1/2$ and $\tilde A^4=1/(4C_1)$, so that there
\be
|\Phi|\sim \frac{1}{\left(4C_1\right)^{1/4}\sqrt{r}}\, .
\ee
\end{itemize}
Evidently, in the case of a massive potential in order to find a non-trivial solution the constraint ($C_1=\lambda^2/(3m^2)$ and $m^2/\lambda<0$) must be satisfied, whereas for the two massless senarios there are always solutions.
A simple check that will be carried out in the next section finds that the constraint is {\it not} satisfied in the case of the massive ABJM model. This will mean that an embedding of the duality into massive ABJM will not be possible and within massless ABJM, only the purely sextic potential will be relevant.

\subsection{Pure sextic potential}

It turns out that in the pure sextic case, $V=C_1|\Phi|^6$, we can solve everything explicitly using some simple considerations. The equation of motion is 
\be
|\Phi|''=3C_1|\Phi|^5\, ,
\ee
and we write it in terms of $v=|\Phi|'$ as 
\be
v\frac{dv}{d|\Phi|}=3C_1|\Phi|^5\, ,
\ee
solved by 
\be
v^2=C_1|\Phi|^6+K_1\Rightarrow |\Phi|'=\pm \sqrt{C_1|\Phi|^6+K_1}\, .\label{eq.phiprime}
\ee
The general solution is then
\be
r+K_2/\sqrt{C_1}=\pm \int\frac{d|\Phi|}{\sqrt{C_1|\Phi|^6+K_1}}\, .
\ee
Note however that if $|\Phi|\sim Ar$ at $r\sim 0$ and $|\Phi|\sim A/r^n$ at $r\sim \infty$ (for positive $n$), there must exist at least one place where $d|\Phi|/dr=0$ in the middle,
or where $v=0$, which implies $C_1|\Phi|^6+K_1=0$, i.e. 
$|\Phi|_{mid}=\left(-K_1/C_1\right)^{\frac{1}{6}}$, which in turn means that\footnote{Note that we can (and in 
general should) glue different branches of the solution at the point where $v=0$.} $K_1/C_1<0$. For the branch connected with $r=0$ however, it is clear from equation (\ref{eq.phiprime}) that having $K_1<0, C_1>0$ would mean that $|\Phi|'$ was imaginary and therefore that $C_1<0$ and $K_1>0$. However, this is inconsistent as we would have a runaway potential with no stable vacuum. 

We must therefore choose $C_1<0$ and $K_1>0$. This choice is, if anything, worse since it implies that the potential is negative definite. In fact even if the vacuum were stable in this case there would be a problem because the solution
\be
|\Phi|'=+\sqrt{K_1-|C_1||\Phi|^6}\,,
\ee
until we reach $|\Phi|_{mid}$, and thereafter
\be
|\Phi|'=-\sqrt{K_1-|C_1||\Phi|^6}\,.
\ee
This means that we would reach $|\Phi|=0$ with nonzero derivative, $|\Phi|'=-\sqrt{K_1}$. Since $|\Phi|\geq 0$, this results in a singularity at this point, as $|\Phi|'$ would jump discontinously. 

In other words, there is {\it no normal smooth solution for the vortex}. This will however not be a problem as the smoothness constraint is not required. We saw that in any case the solution is not valid at $r=0$ itself, so we can ignore the constraint that $|\Phi|=0$ there. 
With a little more thought it is clear that, with $C_1>0$ 
as it should be, the only solution that makes sense (which goes to zero at infinity) is one with $K_1=0$, since if $K_1<0$, $|\Phi|'$ must become imaginary before reaching $r=\infty$, and if $K_1>0$, $|\Phi|'$ must remain finite as $|\Phi|=0$, which means it is again reached before $r=\infty$. Then the 
solution is
\be
\sqrt{C_1}r+K_2=-\int\frac{d|\Phi|}{|\Phi|^3}=\frac{1}{2|\Phi|^2}\, ,
\ee
(we can easily see that the + in front of the integral also doesn't make sense), so that
\be
|\Phi|=\frac{1}{\sqrt{\sqrt{C_1}2r+2K_2}}\, ,
\ee
which has 
\be
|\Phi|'(0)=-\frac{\sqrt{C_1}}{\sqrt{2K_2}}\, ,
\ee
which is finite, but as we said, we must excise and regularize an infinitesimal region around $r=0$. To conclude this section, there is a strong constraint on the form of the sextic potential in the massive case which leads to a vortex solution, whereas for a purely sextic potential, there will be non-smooth solutions which, with excision of the irregular core, will correspond to vortices.

\section{Embedding particle-vortex duality in ABJM}
In order to formulate the particle-vortex duality within ABJM we must be able to find an abelian reduction of the ABJM model which can both be mapped to the path integral in equation (\ref{finalduality}) as well as shown to fulfill the constraint which leads to vortex solutions. We will show below that while the mass-deformed ABJM theory has the appropriate mapping to the self-dual action with non-zero mass, the vortex constraints on the potential are not fulfilled and thus we can only get a self-dual theory with vortices in the massless case. See Appendix \ref{app1} for a brief overview of the ABJM formalism.

\subsection{Constructing a self-dual abelian reduction of ABJM}\label{ABJM2}

For the two bifundamental scalars, $\Phi$ and $\chi$, of ABJM we split the  $N-$dimensional matrix space into two (block-diagonal) $N/2$ dimensional subspaces. In doing so we will be able to use each of the sub-spaces to construct a self-duality under the particle-vortex transformation. In the first subspace, we write the ansatz
\bea
&&A_\mu=a_\mu^{(1)} \mathbf{1}_{N/2\times N/2}\,,\cr
&&\hat A_\mu=\hat a_\mu ^{(1)}\mathbf{1}_{N/2\times N/2}\,,\cr
&&Q^1=\phi G^1_{N/2\times N/2}\,,\cr
&&Q^2=\phi G^2_{N/2\times N/2}\,,\cr
&& R^\a=0\,,\label{embedABJM}
\eea
where $A_\mu$ and $\hat A_\mu$ live in the two gauge groups making up the $U(N)\times U(N)$ gauge symmetry of ABJM and $Q^\alpha$ and $R^\alpha$ with $\alpha=1,2$ are the two (first sub-space) bifundamental scalars. The combination of $Q^\alpha$ and $R^\alpha$, often labeled $N^\alpha$ can be shown with this choice of $R^\alpha$ to vanish while the other combination will be non-zero. The covariant derivative on the scalar $Q^\alpha$ is given by
\be
D_\mu Q^\a=G^\a(\d_\mu \phi+i(a^{(1)}_\mu-\hat a^{(1)}_\mu)\phi)\,,
\ee
where the $N/2\times N/2$ subscript is now left off the $G^\alpha$ for brevity. This leads to the kinetic terms
\be
\Tr[|D_\mu Q^\a|^2]=2\frac{N}{2}\left(\frac{N}{2}-1\right)|\d_\mu+i(a^{(1)}_\mu-\hat a^{(1)}_\mu)\phi|^2\,.
\ee
The second contribution to the mass deformed potential is 
\be
M^\a=\mu Q^\a+\frac{2\pi}{k}(Q^\a Q^\dagger_\b Q^\b-Q^\b Q^\dagger_\b Q^\a)=G^\a\left(\mu \phi+\frac{2\pi}{k}\phi^3\right)\,,
\ee
and thus the full potential
\be
V=\Tr[|M^\a|^2]=\frac{N}{2}\left(\frac{N}{2}-1\right)|\phi|^2\left|\mu+\frac{2\pi}{k}\phi^2\right|^2
=\frac{N}{2}\left(\frac{N}{2}-1\right)|\phi|^2\left(\mu+\frac{2\pi}{k}|\phi|^2\right)^2\,.
\ee
With this ansatz, the Chern-Simons terms reduce to 
\be
\frac{k}{4\pi}\frac{N}{2}\epsilon^{\mu\nu\rho}(a^{(1)}_\mu\d_\nu a^{(1)}_\rho-\hat a^{(1)}_\mu\d_\nu \hat a^{(1)}_\rho)
=\frac{k}{4\pi}\frac{N}{2}\epsilon^{\mu\nu\rho}(a^{(1)}_\mu+\hat a^{(1)}_\mu)\d_\nu(a^{(1)}_\rho-\hat a^{(1)}_\rho)\,,
\ee
so we obtain the first half of the required action, with the additional identification
$\Phi\rightarrow \phi$, $a_\mu\rightarrow a^{(1)}_\mu-\hat a^{(1)}_\mu$ and ${\tilde b}_\mu\rightarrow
a^{(1)}_\mu+\hat a^{(1)}_\mu$. The other half, for $\chi$ and ${\cal A}_\mu$, is obtained from the second $N/2$ subspace, now with  the constraint $\tilde b_\mu=a^{(1)}_\mu+\hat a^{(1)}_\mu=a^{(2)}_\mu+\hat a^{(2)}_\mu$. 

\subsection{Vortex constraints on the ABJM potential}

We are now in a position to construct a duality in this constrained sector of ABJM by mapping the action to a known self-dual action. However, to prove that this is a particle-vortex duality, we first need to show that there is enough freedom in the sextic potential to provide vortex solutions. In the massive case there is a constraint on the potential that $C_1=\lambda^2/3m^2$ in order to have a vortex, which means that we must have
\be
V=\frac{|\phi|^2}{3m^2}(\lambda^2|\phi|^4+3m^2\lambda|\phi|^2+3m^4)=\frac{|\phi|^2}{3m^2}\left[\left(\lambda|\phi|^2+\frac{3m^2}{2}\right)^2+\frac{3m^4}{4}\right]\,,
\ee
in order to have solitons. Clearly, this is not the case for the mass-deformed ABJM model. Therefore at $\mu\neq 0$, the mechanism doesn't work. However at $\mu=0$ (ie. the purely sextic potential), as shown in the previous section, vortex solutions do actually exist. This, along with the field identifications in section (\ref{ABJM2}), suffices to demonstrate that at $\mu=0$ we can construct a reduction of ABJM which exhibits a particle-vortex self-duality.

\subsection{Toward a non-abelian extension}

To close this section we speculate on a possible extension of the particle-vortex duality to non-abelian vortices starting with the observation that with the embedding of the particle vortex duality, we can write it {\em on the reduction ansatz} in the invariant form
\be
\frac{1}{2}\Tr\left[Q^\dagger_\a D^\mu Q^\a-Q^\a(D^\mu Q^\a)^\dagger\right]=\frac{1}{e} \epsilon^{\mu\nu\rho}\d_\nu \Tr(A_\rho +\hat A_\rho)=
\frac{1}{2}\Tr\left[\tilde Q^\dagger_\a \tilde D^\mu \tilde Q^\a-\tilde Q^\a(\tilde D^\mu \tilde Q^\a)^\dagger\right]\,,
\ee
where the trace is taken only on half the matrix space. With the caveat that we have not been able to prove that this holds in general ({\it i.e.} not on the reduction ansatz), it is tempting to think that one can write a nonabelian generalization of the type
\be
\frac{1}{2}\left[Q^\dagger_\a D^\mu Q^\a-Q^\a(D^\mu Q^\a)^\dagger\right]=\frac{1}{e} \epsilon^{\mu\nu\rho}\d_\nu (A_\rho +\hat A_\rho)=
\frac{1}{2}\left[\tilde Q^\dagger_\a \tilde D^\mu \tilde Q^\a-\tilde Q^\a(\tilde D^\mu \tilde Q^\a)^\dagger\right]\,,
\ee
for half the matrix space, and a similar one for the other half. Showing that this is indeed that case in general would go a long way toward generalizing the (self-dual) particle-vortex duality and we leave it as an open problem that we will return to in the future.

\section{Particle vortex duality from Maxwell duality in the bulk, via AdS/CFT}

Having established a framework to understand particle-vortex duality in (at least a reduction of) the ABJM model, we now relate the duality with Maxwell duality in the bulk, via the AdS/CFT correspondence. The partition function in a three-dimensional conformal field theory for a gauge field with a source is generically
(in Euclidean signature)
\be
Z_{CFT}[a_i]=\int {\cal D}\phi e^{- S[\phi]+\int d^3x J^i a_i}\,,\label{zcft}
\ee
($i=1,2,3$), where $\phi$ represents all of the fields in the gauge theory, $J^i$ the $U(1)$ current that couples to the source $a_i$ which is itself the boundary value for the bulk gauge field $A_\mu$.

The corresponding supergravity partition function in the bulk (in Euclidean signature) is given by the bulk Maxwell action in an AdS geometry
\be
Z_{sugra}[a_i]=  e^{-\int d^4x \sqrt{-g}\left[+\frac{1}{4g^2}F_{\mu\nu}^2\right]}\,,
\ee
where $F_{\mu\nu}=\d_\mu A_\nu -\d_\nu A_\mu$ is the bulk gauge field field strength and $\Phi$ corresponds to all dynamical fields in the bulk. We work in the radial gauge $A_z=0$, so $A_i\rightarrow a_i$ on the boundary. We define the four-dimensional Maxwell duality,
\be
\tilde F^{\mu\nu}=\frac{1}{2\sqrt{-g}}\epsilon^{\mu\nu\rho\sigma}F_{\rho\sigma}\,,\label{mxd}
\ee
in terms of which we can rewrite the partition function as 
\be
Z_{sugra}[a_i]=Z_{sugra}[\tilde a_i]= e^{-\int d^4x \sqrt{-g}\left[+\frac{1}{4g^2}\tilde F_{\mu\nu}^2\right]}\,.
\ee
The question is how to relate this to the particle-vortex duality we have already found, in a theory with a known gravity dual. The field theory dual to the self-dual Maxwell theory in the bulk can itself be rewritten, defining a particle-vortex type duality for currents similar to (\ref{currentduality})
\be
J^i=\frac{1}{2}\epsilon^{ijk}\d_j \tilde J_k\,,\label{pvd}
\ee
as
\be
Z_{CFT}[a_i]=\int {\cal D}\phi e^{- S[\phi]+\int d^3x \frac{1}{2}\epsilon^{ijk}(\d_j\tilde J_k) a_i}=
\int {\cal D}\phi e^{-S[\phi]+\int d^3x \tilde J^i(\frac{1}{2}\epsilon^{ijk}\d_j a_k)}\,,
\ee
so that, if 
\be
\tilde a^i=\frac{1}{2}\epsilon^{ijk}\d_j a_k\,,\label{naive}
\ee
it would be written in exactly the form to match $Z_{sugra}[\tilde a_i]$, thus relating the particle-vortex duality (\ref{pvd}) in the CFT with the Maxwell duality (\ref{mxd}) in the bulk.

\subsection{Maxwell duality in $AdS_4$}

Having identified a link between a generic $d+1$-dimensional Maxwell duality and a $d$-dimensional particle-vortex-like duality we turn specifically to the Maxwell duality in $AdS_4$. In Poincar\'{e} coordinates,
\be
ds^2=\frac{-dt^2+dx^2+dy^2+dz^2}{z^2}\,,
\ee
the Maxwell duality (\ref{mxd}) becomes 
\be
\tilde F_{01}=-F_{23};\;\; \tilde F_{23}=-F_{01},...
\ee
i.e. exchanging electric and magnetic components, including in the radial direction.
In the radial gauge $A_3=\tilde A_3=0$ now, we have $F_{23}=-\d_3 A_2$ and $\tilde F_{23}=-\d_3\tilde A_2$, so 
\be
\tilde F_{01}(z=0)=\d_z A_2(z=0);\;\;\;
F_{01}(z=0)=\d_z\tilde A_2 (z=0)\,,
\ee
where $z=0$ is the boundary of $AdS$. Expanding near the boundary
\bea
&&A_i=a_i+z\bar a_i+\frac{z^2}{2}a_i^{(2)}+\frac{z^3}{3!}a_i^{(3)}+...\cr
&&\tilde A_i=\tilde a_i+z\tilde{\bar a}_i+\frac{z^2}{2}\tilde a_i^{(2)}+\frac{z^3}{3!}\tilde a_i^{(3)}+...,
\eea
the above Maxwell duality relations give
\be
\tilde f_{ij}=\frac{1}{2}\epsilon_{ijk}\bar a_k;\;\;\;
f_{ij}=\frac{1}{2}\epsilon_{ijk}\tilde{\bar a}_k\, ,
\ee
where $f_{ij}$ corresponds to the field strength coming only from the leading term in the expansion on the boundary (i.e. $f_{ij}=\partial_ia_j-\partial_ja_i$)
as well as 
\bea
&&\tilde{\bar f}_{ij}=-\frac{1}{2}\epsilon_{ijk}\d_l^2a_k =\frac{1}{2}\epsilon_{ijk}a_k^{(2)}\,,\cr
&&\bar f_{ij}=-\frac{1}{2}\epsilon_{ijk}\d_l^2\tilde a_k=\frac{1}{2}\epsilon_{ijk}\tilde a_k^{(2)}\,,
\eea
etc. Again $\bar f_{ij}=\partial_i\bar a_j-\partial_j\bar a_i$. This result is obtained from two applications of the duality transformations. For the first equality, we first write the duality for $\bar a_i$ in terms
of $f_{ij}$ and then take a derivative. For the second, we look at the order $z$ term in the duality for $\tilde F_{ij}$ vs. $A_k$. Equating the two results gives $\d_l^2 a_k=-a_k^{(2)}$. We will see shortly that this appears from the Maxwell equations.

Normally, in $d\neq 4$, one should be able to give only the $a_i$ as boundary condition, but not $\bar a_i$, the subleading term in the $z$ expansion. In $d=4$ however, as a result of the Maxwell duality, we can specify both $a_i$ and $\bar a_i$, or equivalently, both the source $a_i$ and the source for the Maxwell dual, $\tilde a_i$.
In Poincar\'{e} coordinates the Maxwell equations
\be
\d_\rho [\sqrt{g}g^{\rho\mu}g^{\sigma \nu}\d_\mu]A_\nu -
\d_\rho [\sqrt{g}g^{\rho\mu}g^{\sigma \nu}\d_\nu]A_\mu=0\,,
\ee
and since $g_{\mu\nu}=1/z^2\delta_{\mu\nu}$, this reduces to
\be
\d_\rho \delta^{\rho\mu}\d_\mu A_\nu \delta^{\sigma\nu}-\d_\rho\delta^{\mu\rho}\delta^{\sigma\nu}\d_\nu A_\mu=0\,.
\ee
Notice that all explicit factors of $z$ have disappeared from the equation! This happens only in four dimensions. More generally there will be an extra contribution of $(4-d)/z \d_z A_\sigma$,
which means that in particular, the leading term in this equation is of order $1/z$, namely, for $\sigma=i$, it is $(d-4)\bar a_i/z$, implying that $\bar a_i=0$. In the gauge $A_z=0$, we obtain $\d_i \d_z A_i=0$ for $\sigma=z$.
This constrains $\d_i\bar a_i=0$, $\d_ia_i^{(n)}=0$ for $n\geq 2$, leaving only $\d_i a_i$ possibly nonzero. However, since it is perfectly consistent to set it to zero, we will do so. This is equivalent to the usual radiation gauge with time replaced by $z$, $a_z=0$ and $\d_i a_i=0$. For $\sigma=i$ we obtain 
\be
\d_j^2 A_i+\d_z^2 A_i-\d_i(\d_j A_j)=0\,,
\ee
which when expanded in $z$ (and taking into account the conditions above for $\sigma=z$), results in the system of equations 
\bea
&&\d_j^2 a_i+a_i^{(2)}-\d_i(\d_j a_j)=0\,,\cr
&&\d_j^2\bar a_i+a_i^{(3)}=0\,,\cr
&&\d_j^2a_i^{(n)}+a_i^{(n+2)}=0\,.
\eea
Note that the first relation also implies $\d_i a_i^{(2)}=0$, as it should. Thus, in the radiation gauge for $a_i$ we have $a_i^{(n+2)}=-\d_j^2a_i^{(n)}$. Specifying $a_i$ and $\bar a_i$ (or equivalently, $\tilde a_i$) then completely fixes the solution to the Maxwell equation in $AdS_4$.

Returning to the gauge theory side of the correspondence,  we need to specify $a_i$ and $\tilde a_i$ as sources for the path integral (\ref{zcft}), or exchange $a_i$ with $ \tilde{\bar a}_i$ and 
$\tilde a_i$ with $ \bar a_i$. As claimed earlier, this exchange of $a_i$ with $\tilde{\bar a}_i$ corresponds to a particle-vortex duality exchanging dual currents as in (\ref{pvd}). These currents however, need to be currents of global symmetries that can couple to the gravity dual gauge fields. We need to have two currents, one 
for particles and one for vortices, that can be replaced by their corresponding particle-vortex dual currents.
According to our embedding of particle-vortex duality in ABJM (\ref{embedABJM}), the scalar $\phi$ appears in half of the $U(N)$ space and $\chi$ in the other half. With this ansatz $j^\mu=\hat j^\mu$ from (\ref{jmuhatjmu}) but splits into two currents (for each of the two $N/2$ subspaces) of $\tilde J$ type in 
(\ref{pvd}), $\tilde J^{(1)}_i$ and $\tilde J^{(2)}_i$ that couple to $\bar a_k$ and $\tilde{\bar a}_k$ respectively.

\section{Conclusions}
This article details our exploration of holographic particle-vortex duality. In particular we have foccused on its realization in the ABJM model and a possible relation to Maxwell duality in $AdS_{4}$ via the AdS/CFT correspondence. By combining a path integral version of particle-vortex duality with the Mukhi-Papageorgakis Higgs mechanism we have formulated a symmetric version of the transformation that acts as a self-duality. We then proceeded to show how to embed it as an abelian duality in the $(2+1)-$dimensional, $\mathcal{N}=6$ super Chern-Simons-matter theory that is the ABJM model and speculated on a possible non-abelian extension. Going to the gravity side of the correspondence, Maxwell duality in $AdS_4$ is found to reduce on the 
boundary to a particle-vortex duality acting on two independent gauge field sources $\bar a$ and $\tilde{\bar a}$ 
and their associated currents $\tilde J^{(1)}$ and $\tilde J^{(2)}$.\\

Our primary motivation for this work was two-fold; first we simply wanted to understand if particle-vortex duality is realized in the (mass-deformed) ABJM model with its rich solitonic spectrum and second, we wanted to see if the phenomenological work of \cite{Herzog:2007ij} could be embedded in the concrete setting of the $AdS_{4}\times\mathbb{CP}^{3}$/ABJM correspondence. This work paves the way for both these directions but there remains much to be done. Among the possible extensions of this work are
\begin{itemize}
\item the development of our speculations on a non-abelian version of the particle-vortex duality. To the best of our knowledge the duality has thus far been formulated only of vortices of the conventional Nielsen-Olesen type exhibited by the abelian Higgs model and variants thereof. Vortices, however, come in many different forms and flavors\footnote{Pardon the pun.} such as non-abelian as well as semi-local kinds. It would be of great interest to understand if and how the duality applies to these.
\item an understanding of the manifestation of the full particle-vortex duaity on the gravity side of the correspondence. In particular, having established, in this article, that the duality can actually be embedded into (at least some reduction of) the ABJM model, an important development would be to establish precisely how it acts on states of the type IIA superstring on $AdS_{4}\times \mathbb{CP}^{3}$.
\item the extraction of the phenomenological results for quantum critical transport uncovered in \cite{Herzog:2007ij}.
\item a more complete understanding of how the particle-vortex duality of this article relates to level-rank duality and its generalizations discovered by Kutasov and collaborators in recent years. 
\end{itemize} 
It is quite clear that particle-vortex duality should of great interest to both the holographic condensed matter as well as more formal string theoretic communities and we hope that this article will stimulate further work in this area.

\section{Acknowledgements}
We would like to thank Robert de Mello Koch, Dimitrios Giataganis, Ken Intriligator and Sean Hartnoll for useful conversations at various stages of this work.
The work of HN is supported in part by CNPq grant 301219/2010-9 and FAPESP grant 2013/14152-7. JM acknowledges support from the National Research Foundation (NRF) of South Africa under its IPRR and CPRR programs.
NR is supported by a DAAD-AIMS Scholarship; he wishes to thank the Arnold Sommerfeld Centre for Theoretical Physics and the Max Planck Institute for Physics for hospitality, and the DAAD for a Research Grant for Doctoral Candidates and Young Academics which made his visit to Munich possible.

\appendix

\section{Particle-vortex duality \`{a} la Burgess and Dolan}\label{app1}

In this Appendix we review the duality of \cite{Burgess:2000kj}, ignoring some terms that are not essential to our argument.

\subsection{First derivation}

The starting point is (\ref{thetaaction}), the action of an abelian Higgs system of constant modulus, with an external gauge field $A_\mu$. One also introduces a statistical (Chern-Simons) gauge field\footnote{This field, which arises from the combinatorics of the charged particles, has no dynamical degrees of freedom.} $a_\mu$:
\bea
{\cal L}(\phi,a,A) &=& -\frac{\kappa}{2}[\d_\mu \phi-q_\phi(a_\mu+A_\mu)\phi]^2-\frac{\pi}{2\theta}\epsilon^{\mu\nu\rho}a_\mu\d_\nu a_\rho\cr
&& +{\cal L}_p(\xi,a+A)\;,
\eea
where
\be
{\cal L}_p(\xi,a + A) = \sum_k\left[\frac{m}{2}\dot\xi_k^\mu \dot\xi_{k,\mu} + q_k\dot\xi_k^\mu(a + A)_\mu\right]\delta[x - \xi_k(t)]
\ee
is the particle Lagrangian. Here, $\theta=2\pi n$ for bosons and $\theta=(2n+1)\pi$ for fermions, and $\phi$ is the phase angle of $\Phi=|\Phi|e^{-i\phi}$.

As is usual for dualities in the path integral formulation, we lift this action to a \emph{master action} through the coupling of $\phi$ to a new gauge field ${\cal A}_\mu$ constrained by a Lagrange-multiplier field $b_\mu$ to be pure gauge:
\bea
{\cal L}&=&-\frac{\kappa}{2}[\d_\mu \phi-q_\phi(a_\mu + A_\mu+{\cal A}_\mu)\phi]^2 - \frac{\pi}{2\theta}\epsilon^{\mu\nu\rho}a_\mu\d_\nu a_\rho\cr
&&+{\cal L}_p(\xi,a+A) + \epsilon_{\mu\nu\rho}b_\mu \d_\nu {\cal A}_\rho + \ldots\label{masterbl}
\eea
Indeed, integrating over $b_\mu$, we find $\d_{[\nu}{\cal A}_{\rho]}=0$, and then integrating over ${\cal A}_\mu$ is equivalent to putting it to zero\footnote{Performing the integration over $b_\mu$ produces a functional delta function which enforces the constraint 
$\epsilon^{\mu\nu\rho}\partial_{\nu}{\cal A}_\rho=0$; this, together with the gauge fixing condition, implies that 
integrating over $\mathcal{A}_\mu$ is equivalent to setting $\mathcal{A}_\mu=0$.}. On the other hand, integrating first over $\phi$ instead, 
and then over ${\cal A}$, will lead to a dual action in terms of the Lagrange multiplier $b_\mu$.

To do that, care must be taken about the periodicity of $\phi$ in the presence of vortices for the original complex scalar field $\Phi$.
We have
\be
\phi(\theta+2\pi)=\phi(\theta)+2\pi \sum_aN_a\;,
\ee
where $N_a$ is the \emph{vorticity} or \emph{winding number} of vortex $a$. We then write $\phi=\omega+\varphi$, where $\varphi$ satisfies periodic boundary conditions, $\varphi(\theta+2\pi)=\varphi(\theta)$, and 
$\omega(x)$ is an explicit vortex solution,
\be\label{A5}
\omega(x)=\sum_a N_a \arctan \left(\frac{x^1-y^1_a}{x^2-y^2_a}\right)\equiv \sum_a N_a \theta_a\;,
\ee
where 
\be
\frac{x^1-y_a^1}{x^2-y^2_a}=\tan \theta_a
\ee 
defines the angle of rotation around a particular vortex. In the notation of \cite{Zee} described in the introduction, 
$\omega$ corresponds to $\theta_{\rm vortex}$, and $\varphi$ to $\theta_{\rm smooth}$.

We then define $v_\mu=\d_\mu \omega$, obtaining
\be
v_\mu=\sum_a N_a\frac{1}{1+\tan^2\theta_a}\d_\mu \tan \theta_a=\sum_a N_a \d_\mu \theta_a\;,
\ee
which means that 
\be
\epsilon^{\mu\nu\rho}b_\mu\d_\nu v_\rho = b_\mu \sum_a N_a\ \epsilon^{\mu\nu\rho}\d_\nu\d_\rho \theta_a = 2\pi b_\mu\sum_a N_a\ \dot{y}_a^\mu\ \delta[x - y_a(t)] = 2\pi b_\mu j^\mu(t)\;,
\ee
where 
\be
j^\mu(t) = j^\mu_{\rm vortex}(t)=\sum_a N_a\ \dot{y}_a^\mu\ \delta[x-y_a(t)]\label{vortexcurrent}
\ee
is the vortex current.
Note that $\epsilon^{ij}\d_i\d_j\theta_a = 2\pi \delta^2(x)$, so we can indeed verify the above formula for static $y_a^i(t)=y_a^i$, when $\dot{y}_a^0=1$ and 
the rest are 0, giving
\be
\epsilon^{\mu\nu\rho}\d_\nu \d_\rho\theta_a = \delta^{\mu 0}\epsilon^{ij}\d_i\d_j \theta_a=2\pi \delta^{\mu 0}\delta^2(x-y_a)\;.
\ee
Note now that (\ref{masterbl}) has a gauge invariance
\be
\delta A_\mu =\d_\mu\lambda;\;\;\;\;
\delta\phi=q_\phi \lambda\;,
\ee
which we can gauge-fix by putting $\varphi=0$ (i.e. $\phi=\omega$), thus making the path integration over $\phi$ trivial. We are thus left with only the path integral over ${\cal A}_\mu$ to do, and since $\d_\mu \phi=\d_\mu \omega=v_\mu$, the path integral we need to determine is
\be
\int{\cal D}{\cal A}_\mu \exp\left\{i\int \left[-\frac{\kappa}{2}(v_\mu-q_\phi(a_\mu+A_\mu+{\cal A}_\mu))^2+\epsilon^{\mu\nu\rho}\d_\mu b_\nu {\cal A}_\rho\right]\right\}\;,
\ee
and of course, we still have the particle action and the statistical gauge field part of the action outside the path integral. Then, defining
\be
J^\rho\equiv \epsilon^{\mu\nu\rho}\d_\mu b_\nu\;,
\ee
we get the path integral
\bea
\int {\cal D}{\cal A}_\mu\exp\left[i\int\left( -\frac{\kappa}{2}q_\phi^2\left({\cal A}_\mu + a_\mu + A_\mu - \frac{v_\mu}{q_\phi} - \frac{J_\mu}{kq_\phi^2}\right)^2 - J_\mu \left(a_\mu + A_\mu - \frac{v_\mu}{q_\phi}\right) + \frac{J_\mu^2}{2\kappa q_\phi^2}\right)\right]\cr
={\cal N}\exp\left[i\int\left( -J_\mu \left(a_\mu+A_\mu-\frac{v_\mu}{q_\phi}\right)+\frac{J_\mu^2}{2\kappa q_\phi^2}\right)\right]&&\nonumber
\eea
Given 
\be
\int \frac{J_\mu v^\mu}{q_\phi}=\int \frac{1}{q_\phi}\epsilon^{\mu\nu\rho}\d_\mu b_\nu v_\rho=\int \frac{1}{q_\phi}\epsilon^{\mu\nu\rho}b_\mu \d_\nu v_\rho
=\frac{2\pi}{q_\phi}b_\mu j^{\mu}(t)
\ee
and
\be
J_\mu^2 = 2\delta_{\mu\nu}^{\rho\sigma}\ \d^\mu\ b^\nu \d_\rho b_\sigma = \frac{1}{2}f_{\mu\nu}^{(b)2}\;,
\ee
where $f_{\mu\nu}^{(b)}=\d_\mu b_\nu-\d_\nu b_\mu$, we have as the dual action 
\bea
{\cal L}_{\rm dual}(a,b,A) &=& -\frac{1}{4\kappa q_\phi^2}f_{\mu\nu}^{(b)2} - \epsilon^{\mu\nu\rho}b_\mu \d_\nu (a_\rho+A_\rho) + \sum_a\frac{2\pi}{q_\phi}N_a\ \dot{y}_\mu^a b^\mu\ \delta(x-y_a(t))\cr
&&-\frac{\pi}{2\theta} \epsilon^{\mu\nu\rho}a_\mu \d_\nu a_\rho +{\cal L}_p(\xi, a, A)\;.
\eea
Note that, besides the dualization from the field $\phi$ to the field $b_\mu$, we have also obtained an explicit action for moving vortices, with positions 
$y^a_\mu(t)$, namely $2\pi b_\mu j^\mu_{\rm vortex}(t)$. 
Therefore we explicitly see that the dualization of $\phi$ to $b_\mu$ also exchanges particles with vortices, deserving the name of particle-vortex duality.

\subsection{Second derivation}

We now review a second derivation from \cite{Burgess:2000kj}, which is closer to what we use in the bulk of the paper. We start with an abelian Higgs action where the complex scalar field is coupled to a Chern-Simons gauge field $a$ and an external gauge field $A$, with an arbitrary scalar potential depending only on $|\Phi|$,
\be
S = -\frac{1}{2}\int\left[\left[(i\d_\mu -e\tilde a_\mu )\Phi\right]^\dagger \left[(i\d^\mu -e\tilde a^\mu)\Phi\right] + \frac{\pi e^2}{\theta}\epsilon^{\mu\nu\rho}a_\mu \d_\nu a_\rho\right] + S_{\rm int}\left[|\Phi|^2\right]\;,
\ee
where $\tilde a\equiv a+A$. We rewrite it as 
\be
S = -\frac{1}{2}\int\left[(\d_\mu\Phi)^\dagger(\d^\mu \Phi) + e^2|\Phi|^2\tilde a_\mu \tilde a^\mu - 2\tilde a_\mu j^\mu + \frac{\pi e^2}{\theta}\epsilon^{\mu\nu\rho}a_\mu \d_\nu a_\rho\right] + S_{\rm int}\left[|\Phi|^2\right]\;,
\ee
where the scalar current is
\be
j_\mu =\frac{ie}{2}\left[\Phi^\dagger \d_\mu \Phi -(\d_\mu \Phi^\dagger)\Phi\right]=e|\Phi|^2\d_\mu\theta\;.\label{scalarcurrent}
\ee
Then we split $\Phi$ as before into a vortex part $v$ and a smooth part,
\be
\Phi(\vec{r})=\Phi_0(\vec{r})e^{-i\theta(\vec{r})}v(\vec{r})\;,
\ee
where 
\be
v(\vec{r})=\exp\left[\frac{2\pi i}{q_\phi}\sum_a N_a \arctan \left(\frac{x^1-y^1_a}{x^2-y^2_a}\right)\right]\;.
\ee
Then we have
\bea
S^a[\Phi_0,\theta,a,A] &=& -\frac{1}{2}\int \left[(\d_\mu \Phi_0)^2 + (\d_\mu \theta - iv^*\d_\mu v - e\tilde a_\mu)^2\Phi_0^2\right] + S_{CS}[a] + S_{\rm int}\left[|\Phi|^2\right]\cr \nonumber
\\ \nonumber
&=& -\frac{1}{2}\int\left[(\d_\mu \Phi_0)^2 + e^2\Phi_0^2\tilde a_\mu \tilde a^\mu + \frac{1}{e^2\Phi_0^2}j_\mu j^\mu - 2\tilde a_\mu j^\mu \right] - \frac{\pi e^2}{2\theta}\int\epsilon^{\mu\nu\rho}a_\mu \d_\nu a_\rho\cr
\\
&& + S_{\rm int}[\Phi_0^2]\;,
\eea
where the current is 
\be
j_\mu =e\Phi_0^2(\d_\mu \theta + iv^*\d_\mu v)\;.
\ee
\\
Now we define $\lambda_\mu =\d_\mu \theta$.\footnote{This is the $\d_\mu \varphi$ from the last subsection, whereas $-iv^*\d_\mu v$ is the $\d_\mu \omega$ from there.} 
We substitute the integration over $\theta$ with integration over $\lambda_\mu$, subject to the constraint $\epsilon^{\mu\nu\rho}\d_\nu \lambda_\rho=0$ imposed with a Lagrange multiplier $\tilde b_\mu$, i.e.
\bea
&&\int {\cal D}\theta\exp \left[-\frac{i}{2}\int(\d_\mu \theta + iv^*\d_\mu v - e\tilde a_\mu)^2\Phi_0^2\right]\cr
&=&\int {\cal D}\lambda_\mu {\cal D}\tilde b_\mu \exp\left[-\frac{i}{2}\int(\lambda_\mu + iv^*\d_\mu v - e\tilde a _\mu)^2\Phi_0^2 + \epsilon^{\mu\nu\rho}\tilde b_\mu \d_\nu \lambda_\rho\right]\;.
\eea
Then doing the integral over $\lambda_\mu$ first, we obtain
\bea
S^b[\Phi_0,A,a,\tilde b] &=& \int\left[-\frac{1}{4e^2\Phi_0^2}\tilde f^{(b)}_{\mu\nu}\tilde f^{(b)\mu\nu}+\tilde j^\mu \tilde b_\mu-\epsilon^{\mu\nu\rho}\tilde a_\rho
\d_\nu \tilde b_\rho-\frac{\pi e^2}{2\theta}\epsilon^{\mu\nu\rho}a_\mu \d_\nu a_\rho\right]\cr
&& -\frac{1}{2}\int \d_\mu\Phi_0\d^\mu \Phi_0 + S_{\rm int}'\left[\Phi_0^2\right]\;,
\eea
where as usual, $\tilde f^{(b)}_{\mu\nu}=\d_\mu \tilde b_\nu -\d_\nu \tilde b_\mu$, and 
\be
\tilde j^\mu =\frac{i}{e}\epsilon^{\mu\nu\rho}\d_\nu v^*\d_\rho v
\ee
is the {\em vortex} current.

The duality between the actions $S^a$ and $S^b$ is exactly the same from last subsection, with the kinetic term for $\Phi_0$ and the interaction term being 
spectators, but here it was derived from an abelian Higgs action by path integration. 

Note that the classical solution for $\lambda_\mu$ is 
\be
\d_\mu \theta\equiv \lambda_\mu =-iv^*\d_\mu v+e\tilde a_\mu +\frac{i}{e\Phi_0^2}\epsilon^{\mu\nu\rho}\d_\nu \tilde b_\rho\;, \label{duality2}
\ee
which matches with the duality transform of Zee (from the introduction) with $v=$constant, $\tilde a=0$.

\section{Review of ABJM and its massive deformation}

The ABJM model \cite{Aharony:2008ug} is obtained as the low-energy limit of the theory of $N$ coincident M2-branes in a $\mathbb{C}^4/\mathbb{Z}_k$ background. It is a supersymmetric $\mathcal{N}=6$ $U(N)\times U(N)$ (or $SU(N)\times SU(N)$) Chern-Simons (CS) gauge theory, with bifundamental scalars $Y^I$ and fermions $\psi_I$, $I=1,\ldots,4$ in the fundamental of the $SU(4)_R$ symmetry group, and the two CS gauge fields, $A_\mu$ and $\hat{A}_\mu$, have equal and opposite levels $k$ and $-k$. Its action is 
\begin{align}\nonumber
S &= \int d^3x \left[\frac{k}{4\pi}\epsilon^{\mu\nu\lambda}\Tr\left(A_\mu\partial_\nu A_\lambda + \frac{2i}{3}A_\mu A_\nu A_\lambda - \hat{A}_\mu \partial_\nu \hat{A}_\lambda - \frac{2i}{3}\hat{A}_\mu \hat{A}_\nu \hat{A}_\lambda\right)\right.\\ \nonumber
&\phantom{\int d^3x\frac{k}{4\pi}\ \ }\left. - \Tr\left(D_\mu Y_I^\dagger D^\mu Y^I + i\psi^{I\dagger}\gamma^\mu D_\mu \psi_I\right) + \frac{4\pi^2}{3k^2}\Tr\left(Y^I Y_I^\dagger Y^J Y_J^\dagger Y^K Y_K^\dagger\right.\right.\\ \nonumber
&\phantom{\int d^3x\frac{k}{4\pi}\ \ } \left.\left. + Y_I^\dagger Y^I Y_J^\dagger Y^J Y_K^\dagger Y^K + 4Y^I Y_J^\dagger Y^K Y_I^\dagger Y^J Y_K^\dagger - 6Y^I Y_J^\dagger Y^J Y_I^\dagger Y^K Y_K^\dagger Y^K\right)\right.\\ \nonumber
&\phantom{\int d^3x\frac{k}{4\pi}\ \ }\left. + \frac{2\pi i}{k}\Tr\left(Y_I^\dagger Y^I \psi^J\dagger \psi_J - \psi^{J\dagger}Y^I Y_I^\dagger \psi_J - 2Y_I^\dagger Y^J \psi^{I\dagger}\psi_J + 2\psi^{J\dagger}Y^I Y_J^\dagger \psi_J\right.\right.\\
&\phantom{\int d^3x\frac{k}{4\pi}\ \ }\left.\left. + \epsilon^{IJKL} Y_I^\dagger \psi_J Y_K^\dagger \psi_L - \epsilon_{IJKL}Y^I \psi^{J\dagger}Y^K \psi^{L\dagger}\right)\right]\ . \label{abjm_action}
\end{align}
Here the covariant derivative acts like 
\begin{equation}\nonumber
D_\mu Y^I = \partial_\mu Y^I + i\left(A_\mu Y^I - Y^I\hat{A}_\mu\right)\ .
\end{equation}
The action \eqref{abjm_action} has $SU(4)\times U(1)$ R-symmetry associated with the $\mathcal{N}=6$ supersymmetries.

\subsection{Massive ABJM}

There exists a unique supersymmetry-preserving massive deformation of the model \cite{Hosomichi:2008qk}, parametrised by $\mu$, that breaks the R-symmetry down to $SU(2)\times SU(2)\times U(1)_A\times U(1)_B\times \mathbb{Z}_2$ as a consequence of splitting the scalars as
\begin{equation}\nonumber
 Y^I = (Q^\alpha, R^\alpha), \qquad \alpha = 1,2\ .
\end{equation} 
The $\mathbb{Z}_2$ action interchanges $Q^{\alpha}$ and $R^{\alpha}$, the $SU(2)$ factors act each only on one of the doublets $Q^\alpha$ and $R^\alpha$, 
and the $U(1)_{A}$ symmetry rotates $Q^{\alpha}$ with a charge $+1$ and $R^{\alpha}$ with a charge $-1$. The mass deformation gives mass to the fermions
and  changes the potential of the theory. The bosonic part of the action in the mass deformed case is
\begin{align}\nonumber
\mathcal{L}_{\rm{bosonic}} &= \frac{k}{4\pi}\epsilon^{\mu\nu\lambda}\Tr\left[A_\mu\partial_\nu A_\lambda - \hat{A}_\mu \partial_\nu \hat{A}_\lambda + \frac{2i}{3}\left(A_\mu A_\nu A_\lambda - \hat{A}_\mu \hat{A}_\nu \hat{A}_\lambda\right)\right]\\
&\phantom{\ \ \ } - \Tr|D_\mu Q^\alpha|^2 - \Tr|D_\mu R^\alpha|^2 - V\ .
\label{mass deformed ABJM Lagrangian}
\end{align}
The sextic scalar potential in \eqref{mass deformed ABJM Lagrangian} is
\be\nonumber
V = \Tr\left(|M^\alpha|^2 + |N^\alpha|^2\right)\ ,
\ee
where
\begin{align}\nonumber
M^\alpha &= \mu Q^\alpha + \frac{2\pi}{k}\left(2Q^{[\alpha} Q^\dagger_\beta Q^{\beta]} + R^\beta R^\dagger_\beta Q^\alpha - Q^\alpha R^\dagger_\beta R^\beta + 2Q^\beta R^\dagger_\beta R^\alpha - 2R^\alpha R^\dagger_\beta Q^\beta\right)\ ,\\ \nonumber
N^\alpha &= -\mu R^\alpha + \frac{2\pi}{k}\left(2R^{[\alpha} R^\dagger_\beta R^{\beta]} + Q^\beta  
Q^\dagger_\beta R^\alpha - R^\alpha Q^\dagger_\beta Q^\beta + 2R^\beta Q^\dagger_\beta Q^\alpha - 2Q^\alpha Q^\dagger_\beta R^\beta\right)\ .
\label{mandn}
\end{align}
The equations of motion of the bosonic fields are
\bea
D_\mu D^\mu Q^\alpha &=& \frac{\partial V}{\partial Q^\dagger_\alpha}\ ,\hspace{1cm}
D_\mu D^\mu R^\alpha = \frac{\partial V}{\partial R^\dagger_\alpha}\ ,\nonumber\\
F_{\mu\nu} &=& \frac{2\pi}{k}\epsilon_{\mu\nu\lambda} J^\lambda\ ,\hspace{1cm}
\hat{F}_{\mu\nu} = \frac{2\pi }{k}\epsilon_{\mu\nu\lambda}\hat{J}^\lambda\ ,
\label{EulerLagrangeeqns}
\eea
where $F_{\mu\nu} = \partial_{[\mu} A_{\nu]} + i[A_\mu, A_\nu]$, and the two gauge currents $J^\mu$ and $\hat{J}^\mu$, expressed as
\begin{align}\nonumber
J^\mu &= i\left[Q^\alpha(D^\mu Q^\alpha)^\dagger - (D^\mu Q^\alpha)Q_\alpha^\dagger + R^\alpha(D^\mu R^\alpha)^\dagger - (D^\mu R^\alpha)R_\alpha^\dagger\right]\ ,\\
\hat{J}^\mu &= -i\left[Q_\alpha^\dagger(D^\mu Q^\alpha) - (D^\mu Q^\alpha)^\dagger Q^\alpha + R_\alpha^\dagger(D^\mu R^\alpha) - (D^\mu R^\alpha)^\dagger R^\alpha\right]\ ,\nonumber
\end{align}
are covariantly conserved, i.e. $\nabla_\mu J^\mu = \nabla_\mu\hat J^\mu = 0$. The trace parts of those gauge currents yields two abelian currents $j^\mu$ and $\hat{j}^\mu$ corresponding to the global $U(1)_{A}$ and $U(1)_{B}$ invariances
\begin{equation}
j^\mu = \Tr J^\mu\ ,\qquad \hat{j}^\mu = \Tr\hat{J}^\mu\ ,\label{jmuhatjmu}
\end{equation}
which are ordinarily conserved, i.e. $\partial_\mu j^\mu = \d_\mu \hat j^\mu = 0$. The gauge choice $A_0 = \hat A_0 = 0$ implies that the energy 
density is given by
\begin{equation}\nonumber 
H = \Tr\left[(\partial_0 Q^\alpha)^\dagger(\partial_0 Q^\alpha) + (D_i Q^\alpha)^\dagger(D_i Q^\alpha) + (\partial_0 R^\alpha)^\dagger(\partial_0 R^\alpha) + (D_i R^\alpha)^\dagger(D_i R^\alpha) + V\right]\ .
\label{Hamiltonian}
\end{equation}
Since this is a Chern-Simons theory, varying with respect to $A_0$ and $\hat A_0$ gives the
Gauss law constraints
\begin{align}\nonumber
F_{12} &= \frac{2\pi i}{k}J^0 = \frac{2\pi i}{k}\left[Q^\alpha(\partial^0 Q^\alpha)^\dagger - (\partial^0 Q^\alpha)Q_\alpha^\dagger + R^\alpha(\partial^0 R^\alpha)^\dagger - (\partial^0 R^\alpha)R_\alpha^\dagger\right]\ ,\\
\hat{F}_{12} &= \frac{2\pi i}{k}\hat{J}^{0} = -\frac{2\pi i}{k}\left[Q_\alpha^\dagger(\partial^0 Q^\alpha) - (\partial^0 Q^\alpha)^\dagger Q^\alpha + R_\alpha^\dagger(\partial^0 R^\alpha) - (\partial^0 R^\alpha)^\dagger R^\alpha\right]\ .
\label{gaussconstraint}
\end{align}
Note as an aside that the gauge choice does not uniquely specify the Hamiltonian. Choosing $A_0$ and $\hat A_0$ different from zero introduces 
an extra term in the Hamiltonian, $\epsilon^{\mu\nu\lambda}\Tr[A_\mu A_\nu A_\lambda - \hat A_\mu\hat A_\nu \hat A_\lambda]$. In the abelianisation ansatz of \cite{Mohammed:2012rd}, this vanishes anyway since it is proportional to $\epsilon^{\mu\nu\lambda}a_\mu^{(i)}a_\nu^{(j)}a_\lambda^{(k)}$ and there are only two $a^{(i)}_\mu$'s. So in the abelian case, the Hamiltonian is the same even away from the gauge $A_0=\hat A_0=0$.
 
The mass deformed theory has fuzzy sphere ground states given by\footnote{General vacuum configurations could also be direct sums of these irreducible solutions.} 
\be
R^\alpha = c G^\alpha;\;\;\; Q^\alpha = 0\;\;\;{\rm and}\;\;\;
Q^\dagger_\alpha = c G^\alpha;\;\;\; R^\alpha=0
\ee
where $c\equiv\sqrt{\frac{\mu k}{2\pi}}$ and the matrices $G^\alpha$, $\alpha = 1,2$, satisfy the equations 
\cite{Gomis:2008vc,Terashima:2008sy}
\be
G^\alpha = G^\alpha G^\dagger_\beta G^\beta - G^\beta G^\dagger_\beta G^\alpha.
\ee
In \cite{Nastase:2009ny,Nastase:2010uy}, it was shown that this solution corresponds to a fuzzy 2-sphere, not a 3-sphere as originally thought.

An explicit solution of these equations, which is the unique irreducible one up to a $U(N)\times U(N)$ gauge transformation, is given by 
\bea\label{BPSmatrices}
&& (G^1)_{m,n}    = \sqrt{m - 1} ~ \delta_{m,n}\ , \cr
&& (G^2)_{m,n} = \sqrt{(N - m)} ~ \delta_{m+1,n}\ , \cr
&& (G_1^{\dagger})_{m,n} = \sqrt {m - 1} ~ \delta_{m,n}\ , \cr
&& (G_2^{\dagger})_{m,n} = \sqrt {(N - n)}~ \delta_{n+1,m}\,.
\eea
In particular, $G^1 = G^\dagger_1$. In the case of pure ABJM, instead of a fuzzy sphere ground state, there is a fuzzy funnel {\em BPS solution} 
with $c$ replaced by
\be
c(s)=\sqrt{\frac{k}{4\pi s}}\ . 
\ee
Here $s$ is one of the two spatial coordinates of the ABJM model. The matrices $G^\alpha$ are bifundamental under $U(N)\times U(N)$, 
implying that $G^1G^\dagger_1$ and $G^2 G^\dagger_2$ are in the adjoint of the first 
$U(N)$, and that $G^\dagger_1 G^1$ and $G^\dagger_2 G^2$ are in the adjoint of the second $U(N)$.

\bibliography{PartVort}
\bibliographystyle{utphys}

\end{document}